\documentclass[]{vgtc}                          

\graphicspath{{figures/}{pictures/}{images/}{./}} 

\usepackage{times}                     

\usepackage{tabu}                      
\usepackage{booktabs}                  
\usepackage{lipsum}                    
\usepackage{mwe}                       
\usepackage{comment}

\usepackage{mathptmx}                  
\usepackage{amsmath}
\usepackage{amssymb}
\usepackage{array}
\usepackage{multirow}
\usepackage{verbatim}
\usepackage{graphicx}
\usepackage{xcolor}
\usepackage{url}
\usepackage{enumitem}
\usepackage{pdfpages}

\newcommand{\R}{\mathbb{R}}
\newcommand{\Domain}{\mathcal{M}}

\newcommand{\Sheets}{\mathcal{S}}
\newcommand{\Sheet}{S}

\newcommand{\myparagraph}[1]{\vspace{1mm} \noindent \textbf{#1}}
\newcommand{\ie}{i.e.,\xspace}

\newcommand{\newtext}[1]{\textcolor{black}{#1}\xspace}

\onlineid{1011}

\vgtccategory{Research}

\vgtcinsertpkg

\title{Temporal Tracking of Reeb Space Sheets}

\author{Mohit Sharma\thanks{e-mail: mohit.sharma@liu.se}\\ %
        \scriptsize Link\"oping university %
\and Petar Hristov\thanks{e-mail: petar.georgiev.hristov@liu.se}\\ %
        \scriptsize Link\"oping university %
\and Talha Bin Masood\thanks{e-mail: talha.bin.masood@liu.se}\\ %
        \scriptsize Link\"oping university %
\and Ingrid Hotz\thanks{e-mail: ingrid.hotz@liu.se}\\ %
        \scriptsize Link\"oping university %
}

\teaser{
  \centering
  \includegraphics[width=\linewidth]{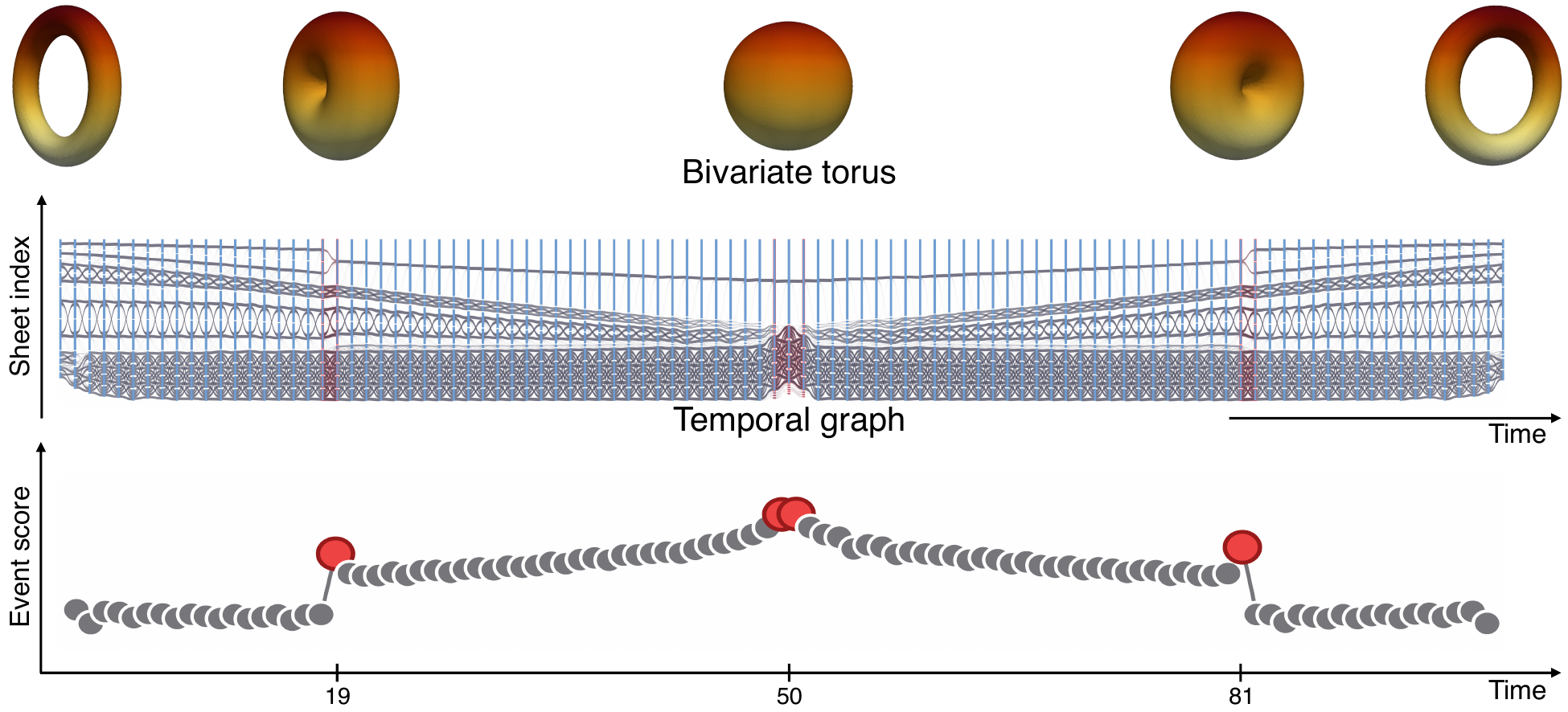}
  \caption{Reeb space sheet tracking for a time-varying bivariate torus dataset. The two scalar fields are an implicit torus function and height. (Top) Isosurfaces of the implicit torus function at a fixed isovalue for selected timesteps from the $100$ timestep sequence, colored by height using a consistent colormap (\includegraphics[width=0.07\textwidth]{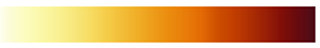}). (Middle) The corresponding sheet tracking temporal graph, which shows near symmetry about the sequence midpoint. (Bottom) Event score plot over adjacent timestep intervals. Higher scores indicate weaker sheet correspondence. The central intervals $49$--$50$ and $50$--$51$ capture the transition from torus to sphere, while the symmetric intervals $18$--$19$ and $81$--$82$ highlight a prominent merge and split event. Nodes and weak links contributing to the highlighted events are shown in red in the temporal graph.}
  \label{fig:teaser}
}

\abstract{ 
Time-varying bivariate fields arise in many scientific applications, where the relationship between two scalar quantities evolves over time. While topological methods such as merge trees provide an effective framework for identifying and tracking features in univariate data, analogous approaches for bivariate fields remain comparatively underexplored. Reeb spaces extend topological analysis to multivariate data by representing fiber connectivity through a collection of interconnected sheets, making these sheets natural candidates for describing bivariate structures. However, establishing temporal correspondences between sheets is challenging due to the structural complexity of Reeb spaces, sensitivity to noise, and the difficulty of defining meaningful similarity measures across timesteps.
We present a framework for tracking Reeb space sheets in time-varying bivariate fields. The method establishes correspondences between sheets in consecutive timesteps using complementary similarity measures defined in the spatial domain and the range space. We evaluate the method on a synthetic torus dataset and two time-varying molecular electronic structure datasets. The results show that Reeb space sheet tracking reveals persistent structures and highlights interesting intervals of temporal change. Overall, the results demonstrate that Reeb space sheets can serve as trackable topological structures and provide a foundation for the visual analysis of time-varying bivariate data.

} 

\keywords{Bivariate field, Reeb space, fiber surface, time-varying data, molecular visualization.}

\begin{document}


\firstsection{Introduction}

\maketitle
Many scientific simulations produce multiple scalar fields over a common spatial domain. When two such fields are analyzed together over time, the data form a time-varying bivariate field. The analysis goal is not only to study each timestep independently, but also to understand how joint structures defined by the two fields evolve across the sequence.

A common strategy for analyzing bivariate fields is to inspect selected isosurfaces or fiber surfaces. Fiber surfaces generalize isosurfaces to bivariate data and provide detailed local views of structures associated with chosen range values. While effective for investigating particular hypotheses, they are less suited for the analysis of feature evolution. Another strategy is to summarize each timestep using a continuous scatterplot (CSP)~\cite{Bachthaler2008}. CSPs generalize discrete scatterplots to continuous spatial fields and capture the distribution of bivariate values. Recent work has used CSP descriptors to analyze the temporal evolution of molecular electronic structure data~\cite{Sharma2025}, where bivariate features are extracted through CSP peeling driven by domain segmentation~\cite{Sharma2021}. Although the resulting descriptors characterize the temporal behavior of peeled CSP layers, they do not explicitly represent the evolution of individual topological features. Consequently, they do not provide correspondences between bivariate features across timesteps, making it difficult to analyze feature persistence, splitting, merging, and reorganization over time.

We address this limitation using Reeb spaces as a topological abstraction of bivariate features. For a bivariate field $F=(f,g):\Domain\rightarrow\R^2$, the Reeb space is obtained by collapsing each connected component of every fiber $F^{-1}(a,b)$ to a single point. Its two-dimensional cells, called sheets, encode both range-space geometry and domain connectedness. A sheet therefore provides an abstraction of a bivariate feature in the domain, and tracking sheets across timesteps captures how such features evolve.

This paper presents a method for tracking Reeb space sheets in time-varying bivariate fields. At each timestep, we compute the Reeb space and retain prominent sheets based on their range-space area. Sheets in adjacent timesteps are compared primarily using range-space shape similarity, which asks whether a sheet preserves a similar relationship between the two scalar fields. Vertex overlap in the domain is used as supporting evidence for spatial consistency and to filter weakly supported correspondences.

The resulting temporal correspondences form a weighted graph that is visualized using a Sankey based layout and linked to sheet renderings, spatial context renderings, and diagnostic plots. The method does not enforce a one-to-one matching, because Reeb space sheets can split, merge, appear, disappear, and change over time. Instead, it retains candidate correspondences that allow the analyst to inspect plausible sheet evolution patterns. This formulation supports the analysis of persistent sheets and important changes in sheet correspondence over time.

The main contributions of this paper are:
\begin{itemize}[leftmargin=*]
  \item A Reeb space sheet tracking workflow for time-varying bivariate fields, where prominent sheets are treated as trackable topological structures.

  \item Complementary correspondence measures for range-space sheet geometry and domain vertex overlap, together with diagnostic summaries.

  \item A linked visual interface that combines Sankey based temporal graphs, event score plots for highlighting potentially interesting timesteps, continuing feature lists, sheet renderings in range space, and spatial context renderings.

 \item Case studies on two molecular datasets, methylvinylketone (MVK) and \textit{cis}-stilbene, and one synthetic torus dataset, demonstrating how Reeb space sheet tracking identifies persistent sheets and important temporal changes in time-varying bivariate data.

\end{itemize}

\begin{figure*}[t]
    \centering
    \includegraphics[width=0.9\textwidth]{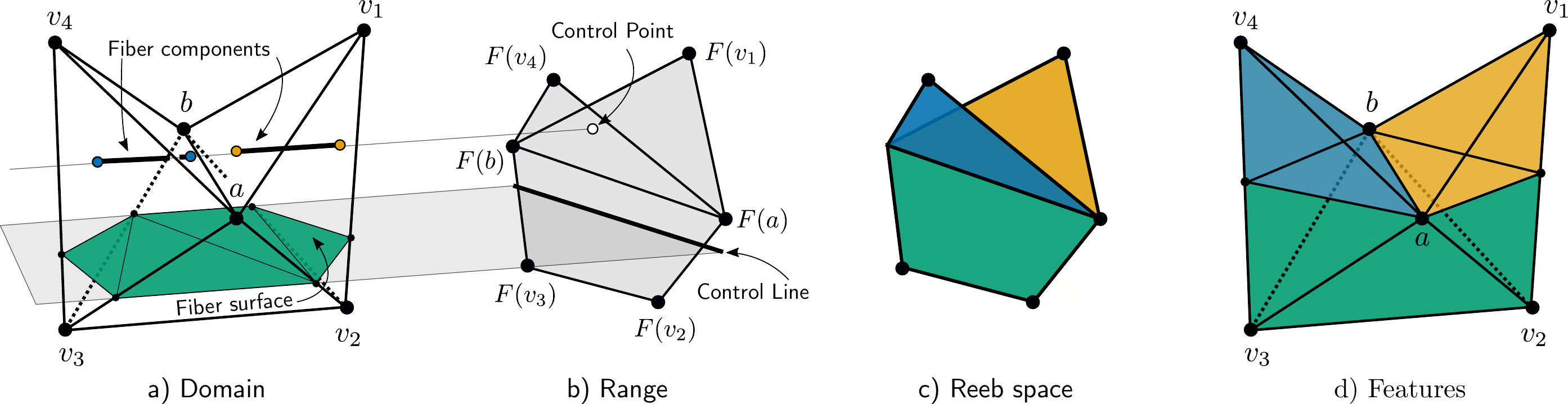}
    \caption{Reeb space and bivariate features. (a) Tetrahedral domain with an example fiber and fiber surface. (b) Range arrangement induced by the images of tetrahedral edges. (c) Three Reeb space sheets obtained by combining faces of the arrangement. (d) Three volume features in the domain, given by the preimages of the sheets.}
    \label{fig:background}
\end{figure*}

\section{Related Work}
Prior work relevant to our method spans three main areas: topological representations for bivariate fields, feature tracking in time-varying data, and visual representations of tracking graphs. We first review Reeb spaces and related topological representations for multivariate fields. We next discuss feature tracking methods with an emphasis on correspondence construction, followed by visualization techniques for exploring temporal tracking graphs.

\myparagraph{Reeb space and bivariate topology.} The Reeb space extends the concept of the Reeb graph to multivariate functions defined over a shared domain~\cite{Edelsbrunner2008}. While a Reeb graph encodes the connectivity of level sets associated with a single scalar function, a Reeb space characterizes how fibers of a multivalued mapping are connected. For bivariate data, the resulting structure is composed of two-dimensional regions that meet along Jacobi edges~\cite{Edelsbrunner2004b}. 
By compactly capturing the topological arrangement of a multifield, Reeb spaces have become a promising tool in multivariate topological analysis. A variety of algorithms have been introduced to compute Reeb spaces for bivariate functions~\cite{Tierny2017,chattopadhyayAlgorithmFastCorrect2024,Hristov2026SingularAA}, as well as for more general settings involving higher dimensional domains and codomains~\cite{Edelsbrunner2008,hristov2025}. 
Ensuring numerical stability and robustness remains a major challenge in these computations~\cite{Hristov2025b}. To address computational complexity and improve scalability, several approximate constructions have been proposed, including the Joint Contour Net~\cite{Carr2014} and Mapper~\cite{Dey2016,Brown2021}, which offer practical alternatives for exploring large and intricate multivariate datasets. Beyond their role as descriptive topological summaries, Reeb spaces and related approximations have also been employed in shape and data comparison tasks through metrics that quantify differences between topological structures~\cite{Ramamurthi2022a,Agarwal2021,Ramamurthi2024}. \newtext{Complementing these deterministic analyses, uncertainty-aware
multivariate visualization includes feature confidence level
sets~\cite{Sane2021FeatureConfidence} and probabilities of fiber positions under
parametric and nonparametric noise models~\cite{Athawale2023FiberUncertainty}.}

\myparagraph{Approaches to feature tracking.} Feature tracking is a central problem in topological data analysis and scientific visualization. Most existing methods focus on univariate data and generally follow one of two strategies: (i) extracting features independently at each time step and subsequently matching them, or (ii) treating time as part of a spatio-temporal domain and extracting features directly in the combined space~\cite{Post2003}. We focus on the first strategy, where topological features are identified at each time step and linked through explicit correspondence mechanisms.

A variety of topological features have been used for tracking~\cite{Yan2021b}, ranging from \emph{critical points} and \emph{region based features} (defined through sublevel or superlevel sets) to intermediate representations such as groupings of critical points based on merge trees (e.g., crown features), which have been applied to cyclone tracking and medical imaging~\cite{Engelke2020,Nilsson2022Cyclone,Rasheed2022}. Other approaches track extremal structures derived from subsets of the \emph{Morse-Smale complex}~\cite{Li2023,Das2024} or full \emph{topological partitionings} of the domain~\cite{Schnorr2020}.

A key challenge is establishing feature correspondences across time. Most methods use pairwise matching between consecutive time steps, while some employ global optimization to determine feature tracks~\cite{Schnorr2020,Saikia2017}. Correspondences are typically based on geometric, attribute, topological, or combined similarity measures.

Early work introduced \emph{Feature Flow Fields} for tracking critical points through derived temporal vector fields~\cite{Weinkauf2011}. This was later reformulated combinatorially using \emph{Morse matching} based on ascending and descending manifolds~\cite{Reininghaus2011}, leading to applications across multiple domains~\cite{Engelke2020,Nilsson2022Cyclone} and a later probabilistic extension~\cite{Nilsson2023}.

Spatial similarity based methods often rely on overlap- or distance based correspondence. First approaches tracked regions defined by thresholds using spatial overlap and geometric attributes such as volume or center of mass~\cite{Samtaney1994,silver1998tracking}. Similar ideas have been applied to space filling decompositions in combustion simulations~\cite{Weber2011,Schnorr2020}. Leveraging the hierarchical multi-scale representation of topological features, overlap correspondence has been extended to track features across scales~\cite{Widanagamaachchi2012,Saikia2017,lukasczyk2017nested}.

More abstract similarity measures include persistence diagram matching~\cite{Cohen-Steiner2010,Soler2018}, merge tree mappings~\cite{Sridharamurthy2020}, and tree edit distance formulations~\cite{Wetzels2022a,Wetzels2023,Wetzels2024ElectronDensity}. Hybrid approaches combine topological and geometric information~\cite{Yan2022b}. Other methods use correspondences based on optical flow~\cite{Valsangkar2019} or optimal transport formulations for tracking merge trees and extremal structures~\cite{Li2023,Li2025,Das2024}.

\myparagraph{Visualization of tracking graphs.} Finally, several works address the problem of improving the quality and usability of tracking results. Proposed visualizations include \emph{augmented tracking graphs}~\cite{Bremer2010}, \emph{nested tracking graphs}~\cite{lukasczyk2017nested,Nilsson2022Cyclone}, and \emph{Sankey diagram} based representations~\cite{Rasheed2024,Evers2026}. Most of these representations allow semantic filtering and are embedded within interactive visualization frameworks. In parallel, visualization techniques have been developed to improve the interpretability of complex tracking graphs, enhancing readability and supporting the analysis of large scale feature evolution~\cite{Kopp2019,Dobler2024}.

\section{Background}
This section introduces the concepts used throughout the paper. We begin with bivariate fields and their domain and range spaces, then review fibers, fiber surfaces, Reeb spaces, and Reeb space sheets. These concepts provide the basis for formulating sheet tracking as a temporal correspondence problem between prominent Reeb space sheets and their corresponding features.

\myparagraph{Bivariate Fields.} Let $\Domain$ be a triangulation of a regular grid volume or, more generally, a 3-manifold.
A \textit{scalar field} over $\Domain$ is a piecewise-linear (PL) map $f:\Domain \to \R$, typically defined on the vertices of $\Domain$ and barycentrically interpolated elsewhere.
A \textit{bivariate field} is a PL map $F=(f,g):\Domain \to \R^2,$
where $f$ and $g$ are scalar fields defined on the same domain.
We refer to $\Domain$ as the \textit{domain}, to $\R^2$ as the \textit{range space} and to $F(\Domain)$ as the \textit{image} of $\Domain$ under $F$.
A bivariate field is \textit{generic} when the image of no three vertices of $\Domain$ are collinear in the plane, and the images of no three edges of $\Domain$ are concurrent in their interiors~\cite{Hristov2025b}.

\myparagraph{Fibers.} The preimage of a point $p=(a,b)\in\R^2$ defines a \emph{fiber}
$F^{-1}(p)=\{x \in \Domain \mid f(x)=a,\ g(x)=b\}$.
The point $p$ is referred to as a \emph{control point}.
Generally, the fibers of $F$ are collections of one-dimensional curves.
Fibers can only change their connectivity at singular edges and vertices~\cite{hristovHypersweepsConvectiveClouds2022}, which generalize critical vertices in the scalar case.
The preimage $F^{-1}(P)$ of a polygon or line $P \subset \R^2$ in the range is a collection of fibers that sweep out a \emph{fiber surface}~\cite{Carr2015}.
We refer to $P$ as a \emph{control polygon or line}.
Fiber surfaces generalize isosurfaces in that they are used to define boundaries of features when visualizing bivariate data.

\myparagraph{Reeb Spaces.} The Reeb space generalizes the Reeb graph from scalar functions to multivariate maps~\cite{Edelsbrunner2008}.
It captures how the connected components of fibers change for all points over the range of $F$.
The Reeb space is the quotient space $\mathcal{R}_F = \Domain/\sim$, where $\sim$ is the equivalence relation on $\Domain$ such that $x\sim x' \text{ if } F(x) = F(x') \text{ and }
x, x' \text{ lie in the same connected component of } F^{-1}(F(x))$.
Practical algorithms for computing the Reeb space include Jacobi fiber surfaces~\cite{Tierny2017b} and the arrange and traverse~\cite{hristov2025, Hristov2026SingularAA} framework.

\myparagraph{Reeb Space Sheets.} 
For a bivariate map over a three-dimensional domain, the Reeb space is represented by a two-dimensional cell complex, whose 2-cells are referred to as \textit{sheets}.
The sheets of the Reeb space are glued along singular vertices and edges where the connectivity of fibers changes, called the Jacobi structure~\cite{Chattopadhyay2014}.
Preimages of sheets in the domain correspond to features captured by the Reeb space, analogous to how the preimages of edges of the Reeb graph and contour tree correspond to features in scalar fields.

\myparagraph{Sheet and Feature Metrics.}
Each sheet of the Reeb space can be described geometrically by its projection onto the range space as a polygon, that may contain holes.  
In the arrange and traverse framework~\cite{hristov2025, Hristov2026SingularAA}, each such polygon is the union of faces of the geometric arrangement of the images of singular edges. 
The area of each sheet is then the sum of the areas of its associated arrangement faces.
The volume of a feature associated with a sheet can be computed exactly using the Jacobi fiber surfaces that bound it or approximated with the number of regular points in that volume (similarly to the scalar case~\cite{Schneider2008}).

\section{Method}

This section describes our method for tracking Reeb space sheets over time. Given a sequence of bivariate volumetric fields, we extract prominent sheets at each timestep and represent each sheet by its range-space footprint, area, and associated domain vertices. We then estimate correspondences between sheets in adjacent timesteps using range and domain similarity measures. These correspondences define a temporal graph, which is summarized using event scores and continuing feature diagnostics and visualized using a Sankey based layout linked to sheet renderings and spatial context renderings.

\subsection{Input and Preprocessing}

The input to our method is a sequence of volumetric datasets ordered in time. Each timestep contains two scalar fields defined on a common spatial domain, forming a bivariate field $F_i=(f_i,g_i)$. We assume that the field pair is consistently defined across the sequence, so that the corresponding scalar quantities are comparable over time.

As a preprocessing step, we compute the Reeb space of each bivariate field and extract its sheets. For each sheet, we store its geometric footprint in the range, its range-space area, and the associated set of regular domain vertices. In some cases, a sheet may have nonzero range-space area but no associated regular domain vertices. We retain such sheets because they may still capture meaningful range-space geometry. The extracted sheet information provides the basis for sheet ranking, visualization, and correspondence estimation.

\subsection{Top Sheet Selection}

To keep the temporal graph visually manageable, we restrict the analysis to the $N$ most significant sheets at each timestep. We rank sheets by their range-space area, as large sheets have been shown to be of particular interest in bivariate data analysis. This criterion is also analogous to common univariate settings, where feature importance is often measured by persistence, i.e., by the size of the scalar value interval over which a feature exists. Let $\Sheets_i^\mathrm{all}$ denote the set of all sheets extracted at timestep $i$. We retain only the $N$ highest ranked sheets:
\[ \Sheets_i = \operatorname{TopN}(\Sheets_i^\mathrm{all}, N) \]
The parameter $N$ is user-defined and can be adjusted interactively to explore the influence of smaller or less prominent sheets on the resulting temporal analysis.

\subsection{Sheet Correspondence}

To capture the temporal evolution of Reeb space sheets, we establish correspondences between sheets in consecutive timesteps. These correspondences support the detection of feature continuations as well as events such as splits, merges, appearances, disappearances, and substantial geometric changes.

Given a time-varying sequence of bivariate fields $F_0, F_1, \ldots, F_T$, let $\Sheets_i$ denote the set of sheets extracted from the Reeb space of $F_i$. For each pair of consecutive timesteps $(i,i+1)$, we connect sheets in $\Sheets_i$ and $\Sheets_{i+1}$ using an active correspondence score $q$. A high score indicates that two sheets are likely to represent the same evolving bivariate feature, or related parts of a feature undergoing a topological transition. The resulting correspondences define a weighted bipartite graph whose vertices are sheets from consecutive timesteps and whose edge weights are given by $q$. \newtext{This correspondence measure is a local, pairwise heuristic rather than a theoretical guarantee. It assumes that adjacent timesteps are sampled densely enough that persistent sheets retain similar range-space footprints.}

\myparagraph{Range-space similarity.}
The range-space correspondence score is based on the geometric similarity of sheets considered as regions in the bivariate range. A widely used measure for comparing geometric regions is the \emph{Jaccard index}, also known as \emph{Intersection over Union} (IoU). It quantifies the overlap between two regions relative to their combined area. For two sheets $\Sheet$ and $\Sheet'$, we define
$$
q_R(\Sheet,\Sheet') =
\frac{|\Sheet \cap \Sheet'|}
{|\Sheet \cup \Sheet'|},
$$
where $|\cdot|$ denotes the area of a region in range space.

The value of $q_R$ lies in the interval $[0,1]$, where $0$ indicates no overlap and $1$ indicates identical regions. Both overlap and relative area are important for comparing sheet footprints in range space. The Jaccard index captures both aspects and is sensitive to changes in position, size, and shape, making it well suited for tracking sheet evolution in range space.

Because sheet footprints may overlap within a single timestep, a sheet $\Sheet\in\Sheets_i$ can have high range-space overlap with multiple sheets in $\Sheets_{i+1}$. The resulting correspondence scores are pairwise similarities between sheets. They are not normalized over the outgoing links and therefore do not generally sum to one.

In addition to this default score, we implemented alternative range-space similarity measures that emphasize specific geometric properties, including area ratio, bounding box IoU, and centroid similarity. These measures are used primarily for diagnostic and evaluation purposes and are not employed as the primary correspondence score in the reported results. Their definitions are provided in the supplement.

\myparagraph{Domain overlap.}
Domain overlap compares the spatial support of sheets across adjacent timesteps. It complements range-space similarity because sheets with similar range-space geometry may correspond to distinct spatial features. We therefore use overlap in the domain as supporting evidence for range-space correspondences.

Let $\Omega_\Sheet$ and $\Omega_{\Sheet'}$ denote the domain regions associated with a source sheet $\Sheet$ and a target sheet $\Sheet'$, respectively. We denote the size of a domain region by $|\cdot|$, which corresponds to volume in the continuous setting. In our discrete implementation, this volume is approximated by the number of associated regular domain vertices.

A natural normalized measure of domain similarity is the Jaccard index,
$$
d_{D\text{-}jac}(\Sheet,\Sheet') =
\frac{|\Omega_\Sheet \cap \Omega_{\Sheet'}|}
{|\Omega_\Sheet \cup \Omega_{\Sheet'}|},
$$
which lies in $[0,1]$ and measures relative overlap. However, this normalization can assign high values to correspondences supported by only a small number of shared vertices when the union of the two supports is also small. Since domain information is used only as supporting evidence for range-space correspondences, rather than as the primary matching criterion, we instead use the raw domain overlap in the results reported in this paper:
$$
O(\Sheet,\Sheet') = |\Omega_\Sheet \cap \Omega_{\Sheet'}|
$$
This score measures the amount of direct spatial support shared by two sheets. It is not source- or target-normalized, \ie it does not measure what fraction of the source persists in the target, or what fraction of the target is inherited from the source. Instead, it emphasizes the absolute amount of shared domain support when filtering range-space correspondences.

The main results use the range-space score $q_R$ for correspondence computation. Raw domain overlap $O$ is used separately as supporting evidence for inspecting and filtering these correspondences. The visual interface also provides an optional domain-active score $q_D$ for exploratory domain-driven analysis, as described in the supplement.

\subsection{Diagnostic Summaries}
The sheet correspondence graph can contain many candidate links, especially for long time sequences. To guide inspection, we compute two diagnostic summaries: event scores for timestep intervals, and continuing feature lifetimes. These diagnostics are used for ranking and filtering during visual exploration. They are heuristic measures and not intended as formal topological event detectors.

\myparagraph{Event scores.} 
To identify potentially interesting temporal transitions, we define event scores that quantify ambiguity in the sheet correspondence graph. Such ambiguities often arise near topological events, including feature splits and merges, where sheets may have multiple plausible continuations or predecessors. Our measures, therefore, highlight timesteps characterized by weak correspondences, missing continuations, or branching correspondence patterns.

\emph{Best outgoing and incoming correspondences.}
Given an active correspondence score $q$ for an adjacent timestep pair $(i,i+1)$, let $N_i=|\Sheets_i|$ denote the number of sheets at time $t_i$. For a source sheet $\Sheet\in\Sheets_i$, we define its best outgoing score as

$$
B^\mathrm{out}(\Sheet)=
\max_{\Sheet'\in\Sheets_{i+1}} q(\Sheet,\Sheet')
$$
This is the strongest available continuation of $\Sheet$ into the next timestep. Similarly, for a target sheet $\Sheet'\in\Sheets_{i+1}$, we define its best incoming score as
$$
B^\mathrm{in}(\Sheet')=
\max_{\Sheet\in\Sheets_i} q(\Sheet,\Sheet')
$$
This is the strongest available predecessor of $\Sheet'$ from the previous timestep. We also compute the mean best outgoing score
$$
\overline{B^\mathrm{out}_i}=
\frac{1}{N_i}
\sum_{\Sheet\in\Sheets_i}
B^\mathrm{out}(\Sheet),
$$
which summarizes the average strength of the best available continuations from timestep $i$ to timestep $i+1$.

\emph{Temporal consistency.} To quantify weak temporal consistency, we introduce a threshold $\theta$ and count sheets with insufficient best matches. For the source timestep, we define
$$
W^\mathrm{out}_i(\theta)=
\left|
\left\{
\Sheet\in\Sheets_i:
B^\mathrm{out}(\Sheet)<\theta
\right\}
\right|
$$
This measures how many sheets lack a strong continuation into the next timestep. Similarly, for the target timestep,
$$
W^\mathrm{in}_i(\theta)=
\left|
\left\{
\Sheet'\in\Sheets_{i+1}:
B^\mathrm{in}(\Sheet')<\theta
\right\}
\right|
$$
which counts sheets without a strong predecessor. Large values of $W^\mathrm{out}_i$ and $
W^\mathrm{in}_i$ indicate time intervals where many sheets have weak continuations or weak predecessors.

\emph{Correspondence ambiguity.} 
We further quantify branching structure by counting potential split and merge configurations. A source sheet is considered a potential split if it has at least two target sheets with scores above the threshold:
$$
P^\mathrm{split}_i(\theta)=
\left|
\left\{
\Sheet\in\Sheets_i:
\left|
\left\{
\Sheet'\in\Sheets_{i+1}:
q(\Sheet,\Sheet')\geq\theta
\right\}
\right|
\geq 2
\right\}
\right|
$$
Similarly, a target sheet is counted as a possible merge if it has at least two source sheets with scores above the threshold:
$$
P^\mathrm{merge}_i(\theta)=
\left|
\left\{
\Sheet'\in\Sheets_{i+1}:
\left|
\left\{
\Sheet\in\Sheets_i:
q(\Sheet,\Sheet')\geq\theta
\right\}
\right|
\geq 2
\right\}
\right|
$$

\emph{Combined event score.}
The \emph{event score} for interval $i\rightarrow i+1$ is defined as
  $$
  E_i(\theta)=
  W^\mathrm{out}_i(\theta)
  +
  W^\mathrm{in}_i(\theta)
  +
  P^\mathrm{split}_i(\theta)
  +
  P^\mathrm{merge}_i(\theta)
  +
  \left(1-\overline{B^\mathrm{out}_i}\right)N_i
  $$
This score ranks timestep intervals by the degree of weak or ambiguous temporal continuity. The first two terms count sheets that lack a strong outgoing continuation or incoming predecessor. The split and merge terms count one-to-many and many-to-one correspondence patterns, which may indicate structural reconfiguration. However, these patterns are less definitive than weak continuation counts. A sheet may have multiple above-threshold matches because of true branching, partial overlap, similar range-space geometry, or domain containment. The final term increases the score when the average best continuation is low, even if some individual sheets remain above threshold. Thus, a large value of $E_i(\theta)$ indicates an interval where temporal continuity is weak, ambiguous, or both.

We additionally evaluated a half-weight variant in which split and merge candidates contribute less than weak continuations or predecessors. Since this did not materially affect the results, we adopt the equal weight formulation for simplicity and interpretability. The score $E_i(\theta)$ is not normalized across datasets and is intended only for ranking temporal intervals within a given dataset. It should not be interpreted as a probability or as an absolute measure of event likelihood.

\begin{figure}[t]
\centering
 \includegraphics[width=\columnwidth]{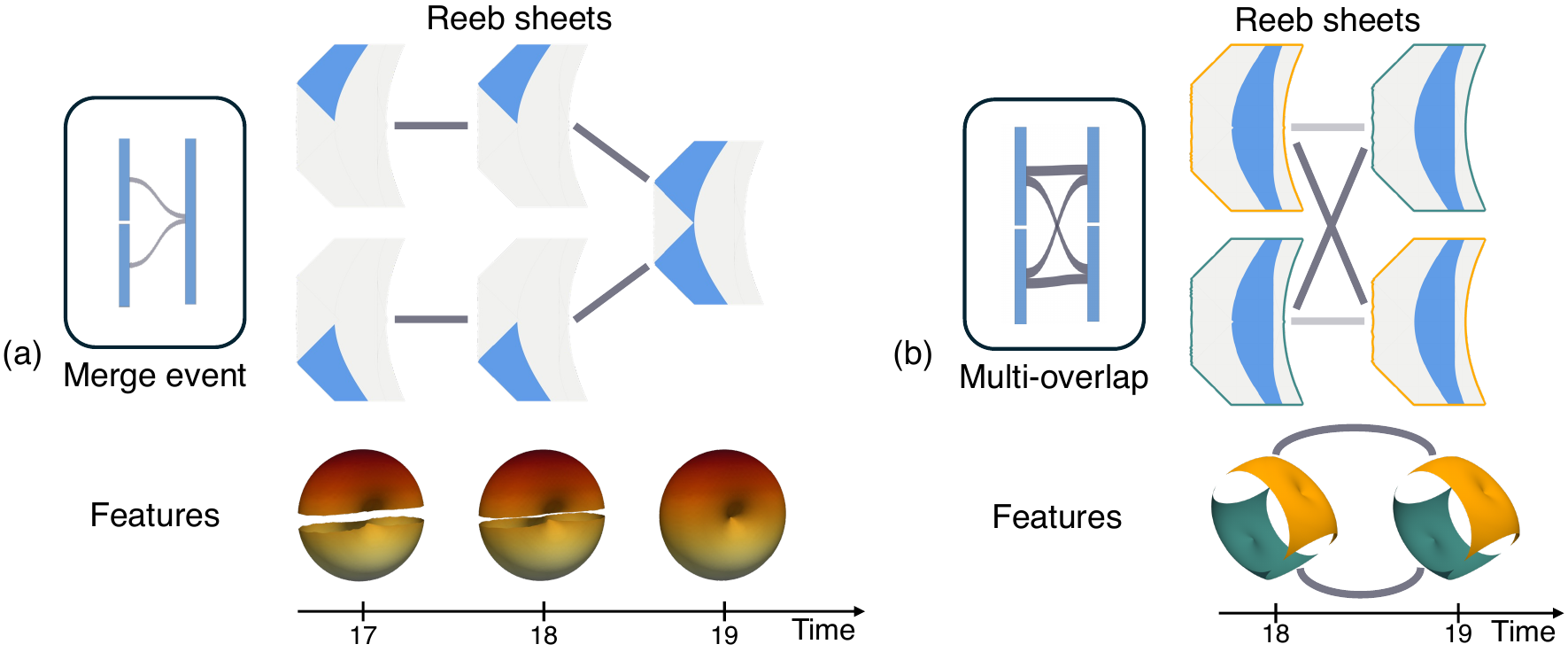}
  \caption{Feature merge and correspondence ambiguity in torus. (a) Interval $18$--$19$, identified by the event score plot, shows two sheets growing, moving closer in the bivariate range, and merging into one. The bottom row shows the corresponding spatial features, rendered as isosurfaces of the torus function restricted to the selected sheets and colored by height. These features also approach each other and merge into a single spatial structure. (b) Correspondence ambiguity occurs when one sheet has strong range-space overlap with multiple sheets in the next timestep. The corresponding spatial features are colored consistently with the sheet boundaries. Domain overlap helps disambiguate these matches by suppressing range-space links with weak domain support, shown with reduced opacity.
 }
  \label{fig:torus-detail}
\end{figure}

\begin{figure*}[t]
  \includegraphics[width=\textwidth]{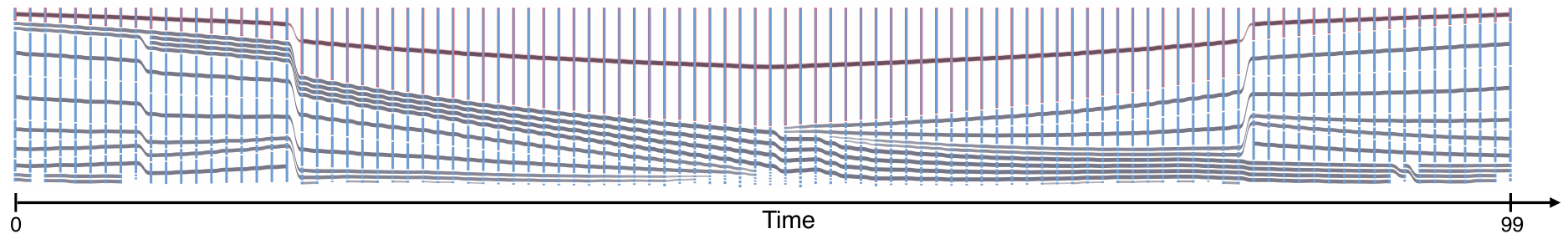}
  \caption{
  Bivariate torus. Simplified temporal graph of the torus dataset after applying domain support filtering. The filtering suppresses correspondences with insufficient domain overlap, reducing link crossings caused by range-space ambiguity. The resulting view emphasizes temporally consistent sheet continuations supported by both range-space similarity and domain overlap. Selecting a node highlights the corresponding longest continuing feature in red.}
  \label{fig:torus-simplified}
\end{figure*}

\emph{Parameter selection.} For the results reported in this paper, event scores are computed using the range-space correspondence score $q_R$. We use $\theta=0.5$ as the main threshold, since it corresponds to substantial overlap between sheet footprints under a common range-space frame. Sensitivity summaries are also exported for multiple $\theta$ values to assess whether highlighted intervals are stable under threshold changes.

\myparagraph{Continuing feature lifetimes.}
To identify persistent sheet behavior, we compute a greedy lifetime diagnostic using the active correspondence score $q$. For a sheet $\Sheet_i\in\Sheets_i$, we select the best target sheet in the next timestep:
$$
\Sheet_{i+1}^{\ast}=
\arg\max_{\Sheet'\in\Sheets_{i+1}}
q(\Sheet_i,\Sheet')
$$
If the best score satisfies
$$
q(\Sheet_i,\Sheet_{i+1}^{\ast})\geq\theta
$$
the track is extended to $\Sheet_{i+1}^{\ast}$, and the same rule is applied again from the next timestep. The lifetime is the number of timesteps visited before the best continuation falls below the threshold or the sequence ends.

For each greedy track, we record its length, mean continuation score, and minimum continuation score. This diagnostic helps identify long lived or stable sheets. However, it does not enforce a global one-to-one assignment, and multiple starting sheets may converge to the same later sheet. Therefore, lifetimes should be interpreted as evidence of local temporal persistence, not as definitive physical feature trajectories.

\begin{figure*}[t]
  \includegraphics[width=\textwidth]{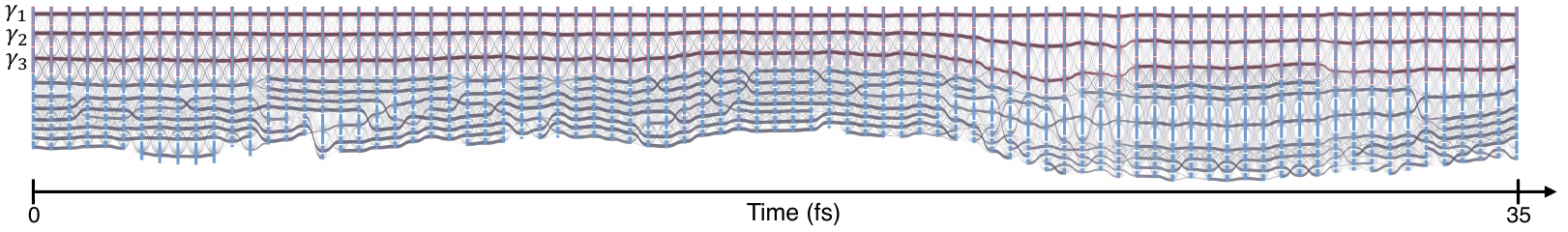}
  \caption{MVK dataset. The temporal graph is computed using the range-space score without domain support filtering. Nodes are sorted by sheet area, and the three most persistent sheets are highlighted in red. Spatial representations of these features are shown in \autoref{fig:MVK-detailed}.}
  \label{fig:MVK-sankey}
\end{figure*}

\begin{figure}[t]
 \includegraphics[width=\columnwidth]{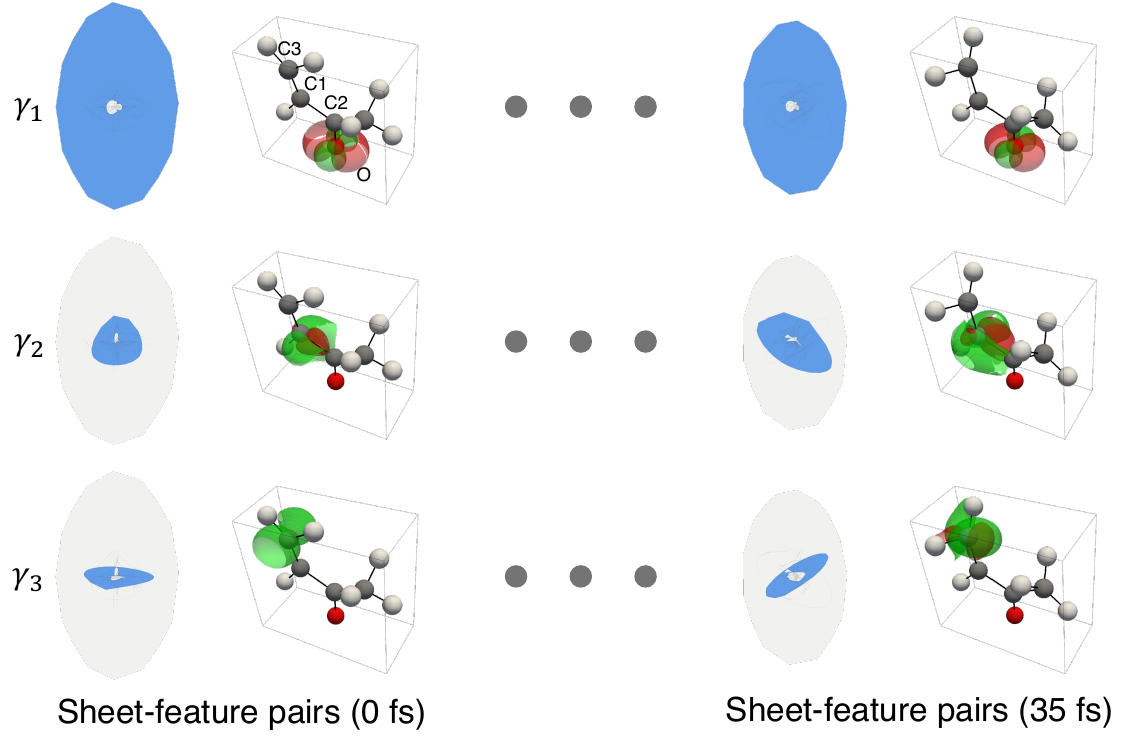}
  \caption{ MVK dataset. Spatial and range-space views of the three tracked features highlighted in \autoref{fig:MVK-sankey}. For each tracked feature, we show the corresponding Reeb space sheet as a footprint in the bivariate range, together with representative sheet-restricted isosurfaces that reveal the associated spatial structure. The tracked feature $\gamma_1$ is localized near the oxygen atom, $\gamma_2$ follows the C1--C2 carbon--carbon bond, and $\gamma_3$ is localized near carbon atom C3.}
  \label{fig:MVK-detailed}
\end{figure}

\begin{figure}[t]
 \includegraphics[width=\columnwidth]{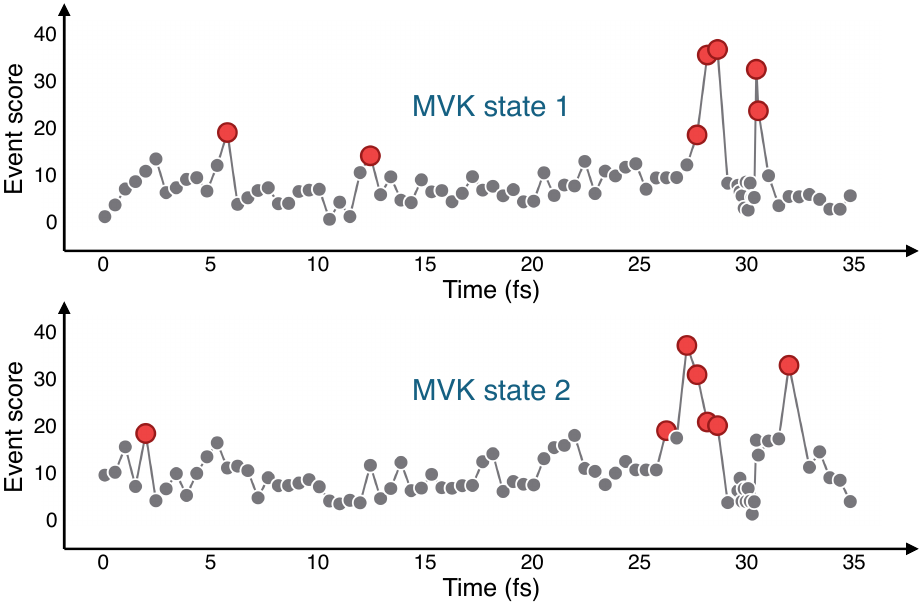}
  \caption{Event score plots for two MVK simulation states. Potentially interesting time intervals are highlighted in red.
}
  \label{fig:MVK-event_score}
\end{figure}

\begin{figure}[t]
 \includegraphics[width=\columnwidth]{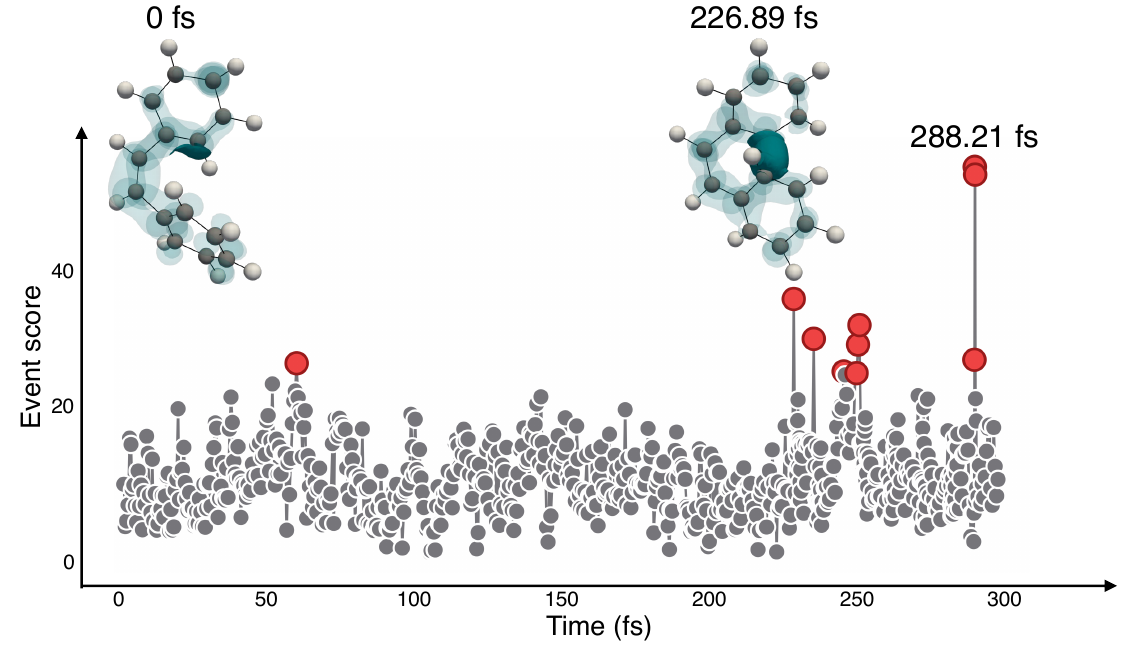}
  \caption{\textit{cis}-stilbene dataset. Event score plot highlighting potentially interesting timestep intervals. The highest scoring interval is analyzed in detail in \autoref{fig:cis_detail}.
}
  \label{fig:event-score-cis}
\end{figure}

\begin{figure}[t]
 \includegraphics[width=\columnwidth]{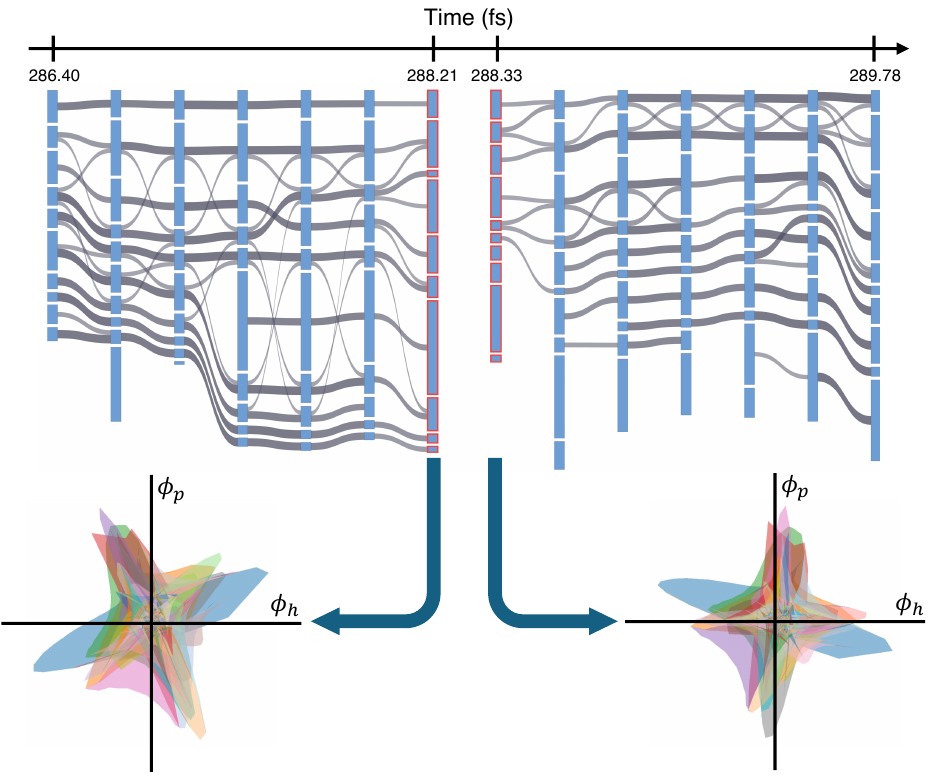}
  \caption{\textit{cis}-stilbene dataset. Local temporal graph around the highest scoring interval, $288.21$--$288.33$~fs. Suppressing links with range-space similarity below $0.41$ removes the connections across the selected interval, indicating weaker temporal continuity than in neighboring intervals. The Reeb spaces shown below the graph exhibit a pronounced rotation, indicating a substantial change in sheet structure that is reflected by the break in the temporal graph.}
  \label{fig:cis_detail}
\end{figure}

\subsection{Sheet and Spatial Context Renderings}

For interactive analysis, we precompute visual representations of the selected sheets and their corresponding spatial context. Each selected sheet is represented by its footprint as a polygonal region in the bivariate range. These range-space footprints are used for both visualization and shape based comparison across timesteps. Each sheet is rendered with a uniform color to emphasize its geometry. The full Reeb space is shown as faint context, while the selected sheet remains visually dominant.

To support meaningful visual comparison over time, all sheet renderings use common range-space bounds and a fixed image size. This ensures that apparent changes in sheet size and position reflect changes in the data, rather than independent rescaling of each rendered image.

For the spatial context shown in this paper, we use sheet-restricted isosurfaces of the individual scalar fields. These renderings are sufficient for our case studies because they reveal the spatial structures associated with the tracked sheets while remaining simple to interpret.

These sheet-restricted isosurfaces should be interpreted as representative spatial context, not as complete descriptions of the sheet's domain support and not as the basis for correspondence computation. More general fiber surfaces can also be computed when an application requires them. In particular, an isosurface of one scalar component can be viewed as a special case of a fiber surface corresponding to a vertical or horizontal control line in the bivariate range.

\subsection{Temporal Graph}

The computed sheet correspondences induce a temporal graph in which nodes represent Reeb space sheets and edges encode correspondence scores between sheets in consecutive timesteps. We visualize this graph using a Sankey based layout, with timesteps arranged from left to right and the top $N$ sheets shown at each timestep. The top sheets are selected by range-space area, providing a stable view of the most visually prominent sheets.

Node height and link thickness encode different quantities. Node height is guided by the number of associated regular domain vertices, using a common scale across timesteps in the dataset. The layout also applies minimum heights for readability, enlarges nodes when necessary so that incident links can be stacked without visual overflow, and may rescale dense columns to keep timesteps readable. Thus, node height should be interpreted as a visual indication of domain support rather than as an exact proportional measure of vertex count.

Link thickness and opacity are encoded independently from node height. They are proportional to the active correspondence score, normalized by the maximum observed value of that score in the dataset. For the results reported in this paper, the active score is the range-space correspondence score $q_R$. This normalization improves readability within a chosen score, but does not make different scores directly comparable.

Unlike conventional Sankey diagrams, our graph does not represent a conserved flow. A sheet may have several strong outgoing or incoming correspondences because sheet footprints can overlap substantially in range space and multiple target sheets may have similar range-space geometry. Consequently, the sum of incoming or outgoing link widths is not expected to equal the node height. This distinction is especially important in the range-space temporal graph: node height indicates domain vertex support, whereas link thickness measures range-space similarity. Thus, a sheet can appear as a large node because it contains many domain vertices, while its links remain thin if its range-space footprint changes, shifts, splits, or only partially overlaps with sheets in the next timestep.

To facilitate exploration, the visualization provides filtering and layout options. Links can be thresholded by the active correspondence score. In addition, support filters can retain the outgoing, incoming, or bidirectional links with the strongest auxiliary support. For the range-space temporal graphs used in this paper, this auxiliary support is the domain overlap $O(\Sheet,\Sheet')$, which measures how many associated domain vertices are shared by the two sheets. Links with insufficient support may remain visible with reduced opacity or be removed entirely.

The temporal graph analyses support exploration by ranking timestep intervals using the event score $E_i(\theta)$, following continuing features through best links above the selected threshold $\theta$, and repeating these summaries over different $\theta$ values to identify intervals whose rankings remain stable across thresholds. The analysis summary plots are linked to the temporal graph. Selecting a node in a plot highlights the corresponding timesteps, nodes, and links in the graph.

Selections in the temporal graph open linked detail views. A node view shows sheet metadata, the rendered sheet footprint, and available spatial context images. A link view shows the source and target sheets side by side. Image views support linked pan and zoom so that corresponding sheet or spatial context renderings can be compared at the same scale. The supplement provides a complete control-level description.

\section{Results}

We evaluate our Reeb space sheet tracking framework on one controlled synthetic dataset and two molecular electronic structure datasets. The synthetic torus dataset contains $100$ timesteps and provides a controlled setting for testing whether the method captures known transitions in sheet structure. The molecular datasets are obtained from photoinduced molecular dynamics simulations and include methylvinylketone (MVK) with $83$ timesteps and \textit{cis}-stilbene with $704$ timesteps.

We provide an \mbox{\href{https://reebflowvisualizer.github.io}{online prototype}} of our visual interface for reader exploration. The prototype includes the torus dataset and two MVK electronic states, which can be selected from the dataset dropdown as \emph{torus}, \emph{MVK s1}, and \emph{MVK s2}.

The molecular case studies follow the setting of previous bivariate analyses of electronic structure evolution~\cite{Sharma2025}. In these data, pairs of scalar fields represent the hole ($\phi_h$) and particle ($\phi_p$) natural transition orbitals (NTOs)~\cite{martin_nto_2023} over time. The two NTO fields describe where electronic density is depleted and where it is accumulated for a given excited state relative to the ground state. We treat these fields as a time-varying bivariate field over evolving nuclear geometries. The goal is to support feature level analysis of this evolution. Our analysis focuses on prominent Reeb space sheets extracted at each timestep, their temporal correspondences, and diagnostic summaries that identify persistent features and intervals of possible structural change.

\newtext{These case studies are complementary to prior CSP/image-moment analyses of the same time-varying fields~\cite{Sharma2025}. Our focus is on feature-level correspondences between Reeb space sheets, persistent sheet tracks, and event intervals that can be inspected through linked range-space and spatial views.}
  
For all datasets, we compute the top $N=20$ sheets at each timestep. The interface displays the top $10$ sheets by default for readability, but this can be increased up to $20$. In the following results, we use the range-space correspondence score $q_R$ as the active score for sheet tracking. Thus, the reported event scores primarily reflect changes in sheet footprints in the bivariate range, while domain overlap is used as auxiliary support for inspecting and filtering correspondences.

\subsection{Bivariate torus}
Our first dataset is a synthetic bivariate torus, defined by a time-varying implicit torus function and the height function. For an intuitive overview, \mbox{\autoref{fig:teaser}~(top)} shows isosurfaces of the torus function at a fixed isovalue for selected timesteps. Around timestep $50$, the torus deforms into a sphere before transforming back into a torus with a different orientation.

We process this dataset using our Reeb space sheet tracking method. As expected, the resulting temporal graph is nearly symmetric around timestep $50$, reflecting the symmetric deformation of the torus. The event score plot exhibits the same symmetry. The selected intervals $18$--$19$ and $81$--$82$ capture two major sheets merging into a single sheet and the resulting sheet later splitting back into two. In the temporal graph, the nodes and links highlighted in red indicate the sheets and correspondences that contribute to the event score for the selected interval, i.e., those involved in weak or ambiguous continuations.

\autoref{fig:torus-detail}(a) shows the merge event in detail. The two sheets increase in range-space area and move closer in the bivariate range at timesteps $17$ and $18$, before merging at timestep $19$. The bottom row shows the corresponding spatial structures, obtained by extracting isosurfaces of the torus function restricted to the individual sheets at a consistent isovalue and coloring them by height. These views show that the two spatial features grow and approach each other at timesteps $17$ and $18$, and merge into a single feature at timestep $19$.

We also observe several crossing links in the temporal graph. An inset from the interval $18$--$19$ is shown in \autoref{fig:torus-detail}(b). These crossings occur because sheets supported on different regions of the domain can have highly similar footprints in range space. As a result, a single sheet can have several strong range-space correspondences, producing visually dense link patterns. As shown in \autoref{fig:torus-detail}(b), both sheets at timestep $18$ have strong links to both sheets at timestep $19$ because of their high range-space similarity. However, the corresponding spatial features, shown using the same colors as the sheet boundaries, exhibit clearer one-to-one support in the domain. This domain support can therefore be used to disambiguate the range-space correspondences. After applying this support, the lighter range-space links shown in \autoref{fig:torus-detail}(b) are suppressed. The visual interface supports this operation through domain support filtering. This filtering either reduces the opacity of, or removes, range-space correspondences that are not sufficiently supported by domain overlap, depending on the user-selected transparency setting. \autoref{fig:torus-simplified} shows the simplified visualization, emphasizing continuations supported by both range-space similarity and domain overlap. One continuing feature is highlighted in red.

The simplified temporal graph is less visually symmetric than the unsimplified one shown in \autoref{fig:teaser}. This difference is caused by the node ordering strategy: the layout orders nodes to minimize weighted link crossings, and this optimization can choose an ordering that reduces crossings rather than preserving the visual symmetry of the dataset. The prototype supports alternative node ordering options, including ordering by sheet area and by associated domain vertex count. These orderings can help preserve a more symmetric view in both simplified and unsimplified visualizations.

\subsection{Methylvinylketone \newtext{(MVK)}}

\newtext{For the MVK dataset, we evaluate whether Reeb space sheet tracking recovers coherent molecular regions over time and identifies intervals reported in earlier analyses~\cite{Sharma2025}.} \autoref{fig:MVK-sankey} shows the complete temporal graph for all $83$ timesteps, generated using the range-space correspondence score $q_R$. Nodes are sorted by sheet area.

We first examine the three longest continuing features, labeled $\gamma_1$, $\gamma_2$, and $\gamma_3$ in the temporal graph. Each feature persists throughout the full sequence of $83$ timesteps and remains visually prominent in the unsimplified graph, even without domain support filtering. \autoref{fig:MVK-detailed} shows the corresponding sheets and representative spatial context renderings at timesteps $0$ and $35$. The spatial context renderings show sheet-restricted isosurfaces of the individual NTO fields at representative field values of magnitude $0.07$, with hole NTO isosurfaces shown in green and particle NTO isosurfaces shown in red. These views reveal coherent spatial structures associated with the tracks and highlight that they are followed consistently from the beginning to the end of the sequence. $\gamma_1$ tracks a region close to the oxygen atom, which is notable because changes near the oxygen region have also been reported as important in previous studies of this molecule~\cite{Sharma2025}. $\gamma_2$ tracks the C1--C2 carbon--carbon bond, while $\gamma_3$ captures the evolution around carbon atom C3. These examples show that the method can track consistent Reeb space sheets and their corresponding spatial features over time.

Next, we examine the event score plots for MVK state $1$ and state $2$, shown in \autoref{fig:MVK-event_score}. Both plots contain several peaks, with the top $7$ peaks highlighted in red. These peaks indicate prominent activity between $27$ fs and $32$ fs, along with additional activity near the initial timesteps. The highlighted intervals agree well between the two states and are consistent with intervals identified as important in previous work~\cite{Florian2024,Sharma2025}. In particular, the interval around $27$--$30$ fs is detected across all tested threshold values $\theta$, suggesting that it is a robustly highlighted interval. Detailed range sensitivity tables are provided in the supplement. Additional feature evolution for MVK state $2$ can be explored through the prototype.

\subsection{\textit{cis}-stilbene}

The \textit{cis}-stilbene dataset consists of $704$ timesteps. Analyzing the full temporal graph is challenging because of the length of the sequence. We therefore first use the event score plot to identify intervals of interest. As shown in \autoref{fig:event-score-cis}, the ten highest scoring intervals are highlighted in red.  These intervals overlap with chemically relevant periods reported in previous work~\cite{Sharma2025}, but chemical interpretation requires inspection of the linked spatial views. In \autoref{fig:event-score-cis}, representative isosurfaces at magnitude $0.05$ are shown in green. The isosurface segment between the rings is rendered opaque, while the remaining parts are shown transparently. This connecting segment is visible at $226.89$~fs but absent at $0$~fs, indicating formation of the inter-ring bond.

We further analyze the highest scoring interval, which occurs between $288.21$~fs and $288.33$~fs. This interval is also identified as prominent for most tested threshold values $\theta$, indicating that it is stable under threshold variation. To inspect this transition, we extract a local temporal graph around the interval and retain only links with range-space similarity of at least $0.41$, as shown in \autoref{fig:cis_detail}. This filtering disconnects the temporal graph at the selected interval, indicating that most correspondences across the interval are weaker than those in neighboring intervals. The discontinuity suggests that the bivariate field undergoes a substantial change during this transition.

To investigate this change further, we show the Reeb spaces at the two timesteps. The two Reeb spaces exhibit a visually significant change, including an apparent rotation of nearly $90^\circ$ in range space. This change is reflected in the temporal graph by the weak range-space correspondences across the interval.

\section{Implementation and performance}
Our implementation combines offline preprocessing with an interactive browser prototype. Reeb spaces and sheet geometries are computed offline using the arrange and traverse algorithms~\cite{hristov2025,Hristov2026SingularAA}. The extracted sheets and associated domain vertices are then used to compute range-space correspondences, domain overlap, event scores, and continuing feature summaries. The resulting sheet, link, and analysis data are stored as JSON artifacts, together with precomputed sheet images and spatial context renderings.

The browser interface does not recompute Reeb spaces or sheet correspondences. It loads the precomputed JSON and image artifacts and supports interactive layout updates, thresholding, opacity changes, range selection, highlighting, and linked detail views. Filtering operations therefore act on the already loaded temporal graph.

\newtext{For a sequence $F_0,\ldots,F_T$, there are $T$ adjacent timestep pairs. 
After selecting the top sheets, $|\Sheets_i| \leq N$ at each timestep $i$. Let $C_R$ denote the cost of evaluating the range-space score $q_R$ for a pair of sheets. 
Excluding Reeb space computation and top sheet selection, correspondence computation examines $ \sum_{i=0}^{T-1} |\Sheets_i||\Sheets_{i+1}| \leq TN^2$ candidate pairs, resulting in a time complexity of $O(TN^2C_R)$. The temporal graph contains $O(TN)$ nodes and at most $O(TN^2)$ links. 
Computing event scores, applying filters, and determining feature lifetimes require scanning these links and therefore take $O(TN^2)$ time in the worst case. 
If domain overlap scores are evaluated for all candidate pairs, their cost is additionally $O(TN^2C_D)$, where $C_D$ is the cost of intersecting the domain vertex sets associated with a pair of sheets.
}

To provide a rough indication of computation cost, we measured the core pipeline on three timestep subsets, retaining the top $20$ sheets per timestep. The measurements used an AMD Ryzen Threadripper 7960X workstation with $24$ physical cores, $48$ hardware threads, and $61$~GiB RAM. The MVK subset contained about $297$K vertices and $1.39$M cells, with $63.3$~MiB of input VTU data. Reeb space extraction and sheet geometry export took about $5$ minutes, and matching sheets in range space took about $2$ minutes. The torus subset contained about $398$K vertices and $1.88$M cells, with $21.9$~MiB of input VTU data. Reeb space extraction and sheet geometry export took about $26$ minutes, while matching sheets in range space took less than $1$ minute.

The difference in Reeb space computation time is not explained by input size alone. The Singular Arrange and Traverse algorithm~\cite{Hristov2026SingularAA} depends on the singular edge structure of the bivariate PL map and on the induced arrangement and traversal. In our diagnostic count for the first timestep, MVK had about $4$K singular edges, whereas torus had about $7$K. This difference, together with arrangement and sheet geometry export costs, partly explains why torus required more preprocessing time despite a mesh size of the same order. The remaining stages, including JSON conversion, domain overlap attachment, event and feature analysis, and viewer data export, each took less than one second in these tests.

Fiber surface or sheet-restricted isosurface computation and rendering are optional linked detail stages and are not used for correspondence computation. For the three timestep MVK subset, computing selected spatial context surfaces took about $1.4$ minutes, and rendering the corresponding images took about $8$ seconds.

For the full datasets, per timestep Reeb space logs indicate the cost of the dominant preprocessing step. Summing these independent logs gives an equivalent serial compute time of about $10.5$ hours for MVK state 1, $10.0$ hours for MVK state 2, $66$ hours for torus, and $636$ hours for \textit{cis}-stilbene. These values are not measured serial runs. They report the total amount of computation accumulated over all timesteps. Since Reeb space extraction is independent across timesteps, the elapsed time can be reduced substantially by running multiple timesteps in parallel.

The main interaction cost is rebuilding the visible temporal graph layout, especially with weighted crossing-reduced ordering. This cost depends on the number of visible nodes and links, not on the original volume size. The supplement reports interaction performance for full sequence views. MVK and torus remain interactive for the full top $20$ view, while for \textit{cis}-stilbene the top $10$ view remains usable and the full top $20$ view is slower because of the larger number of visible links.

\section{Conclusion}
We presented a workflow for tracking Reeb space sheets in time-varying bivariate fields. The method treats linked sheets as temporally trackable features and builds a correspondence graph from range-space sheet similarity, with domain vertex overlap used as auxiliary support. This is deliberately not a conserved flow model: Reeb space sheets can split, merge, overlap in range space, and reorganize in the spatial domain, so retaining multiple candidate correspondences is important for visual analysis.

The results show that Reeb space sheets can be tracked meaningfully over long sequences despite structural instability in Reeb space computation. On the controlled torus sequence, the method recovers the expected symmetry and highlights central transitions, merges, and splits. On the MVK datasets, it identifies sheets that persist across the full sequence and correspond to interpretable molecular regions such as the oxygen region and carbon--carbon bond structure. On the \textit{cis}-stilbene sequence, event score summaries enable inspection of more than $700$ timesteps and isolate intervals where the bivariate field changes substantially. These examples suggest that Reeb space sheets are not only descriptive structures at individual timesteps. When linked across time, they provide useful temporal features for tracking persistent and changing structures.

A central design choice is to use range-space similarity as the primary tracking criterion and domain overlap as supporting evidence. Matching sheets in range space asks whether the bivariate relationship represented by a sheet persists over time. In contrast, domain overlap asks whether the same regular vertices continue to support a sheet. Pure domain overlap tracking can therefore produce unintuitive temporal graphs: domain vertices may scatter across several sheets, or a sheet may match an unexpected target when most of its vertices begin to participate in another sheet at the next timestep. In such cases, domain overlap reveals spatial support transfer but may not preserve the visual identity of the range-space feature. We therefore use domain support to disambiguate plausible range-space matches rather than replacing range-space tracking with tracking based only on domain overlap. Complementary domain-active views, including domain overlap graphs with range support, are discussed in the supplement. These views answer different questions and are less direct for tracking sheets by range-space similarity.

The method has several limitations. First, we track only the top $N$ sheets by range-space area, so smaller but meaningful sheets may be omitted unless the user increases the visible sheet count or reruns the pipeline with a larger $N$. Second, the default range-space IoU uses a fixed raster grid, which is efficient for large batches but can be sensitive for very small or thin sheets. Third, domain overlap scores depend on mesh sampling and on the assignment of regular vertices to sheets. Some sheets have nonzero range-space area but zero associated regular vertices, making them meaningful in the range view but weak or absent in domain support analyses. Fourth, the continuing feature diagnostic is greedy and local. It supports exploration but does not provide a globally optimized assignment of feature tracks. Finally, the workflow still requires substantial preprocessing, especially for Reeb space extraction and image generation, and is therefore best suited to offline preprocessing followed by interactive visual analysis. \newtext{We have not evaluated it on terabyte- or petabyte-scale data; scalable Reeb-space computation is outside the scope of this paper, and once sheets have been extracted, the cost of our correspondence and visualization stages depends primarily on the number of timesteps, retained sheets, and displayed links rather than directly on the raw volume size.}

Future work should investigate more stable sheet descriptors, tracking strategies that retain multiple plausible sheet continuations over time, adaptive sampling for range-space overlap, and tighter coupling between Reeb space sheets and spatial feature volumes. Another important direction is to study domain support transfer explicitly, since such transfers may reveal meaningful physical or chemical reorganization even when the range-space feature identity remains visually stable.

\acknowledgments{This work is partially supported by the Wallenberg AI, Autonomous Systems and Software Program (WASP), funded by the Knut and Alice Wallenberg Foundation; the Swedish Research Council (VR) under Grant 2023-04806; the Swedish e-Science Research Centre (SeRC); and the ELLIIT environment for strategic research in Sweden. The chemistry datasets used in this work are provided by Nanna Holmgaard List from KTH Royal Institute of Technology, Stockholm.}

\newpage

\bibliographystyle{abbrv-doi}

\bibliography{bib}

@online{chattopadhyayAlgorithmFastCorrect2024,
	author = {Chattopadhyay, Amit and Ramamurthi, Yashwanth and Saeki, Osamu},
	date = {2024},
	eprint = {2403.06564},
	eprintclass = {cs},
	eprinttype = {arXiv},
	pubstate = {prepublished},
	title = {An {{Algorithm}} for {{Fast}} and {{Correct Computation}} of {{Reeb Spaces}} for {{PL Bivariate Fields}}},
	urldate = {2025-08-14}}

@article{hristov2025,
	author = {Hristov, Petar and Sakurai, Daisuke and Carr, Hamish and Hotz, Ingrid and Masood, Talha Bin},
	date = {2025},
	doi = {10.1111/cgf.70206},
	issn = {1467-8659},
	journaltitle = {Computer Graphics Forum},
	publisher = {{The Eurographics Association and John Wiley \& Sons Ltd.}},
	title = {Arrange and Traverse Algorithm for Computation of Reeb Spaces of Piecewise Linear Maps}}

@inproceedings{Hristov2026SingularAA,
	address = {Dagstuhl, Germany},
	author = {Hristov, Petar and Hotz, Ingrid and Masood, Talha Bin},
	booktitle = {42nd International Symposium on Computational Geometry (SoCG 2026)},
	doi = {10.4230/LIPIcs.SoCG.2026.57},
	editor = {Ahn, Hee-Kap and Hoffmann, Michael and Nayyeri, Amir},
	pages = {57:1--57:17},
	publisher = {Schloss Dagstuhl -- Leibniz-Zentrum f{\"u}r Informatik},
	series = {Leibniz International Proceedings in Informatics (LIPIcs)},
	title = {{Singular Arrange and Traverse Algorithm for Computing Reeb Spaces of Bivariate PL Maps}},
	volume = {367},
	year = {2026}}

@inproceedings{Hristov2025b,
	author = {Hristov, Petar and Hotz, Ingrid and Masood, Talha Bin},
	booktitle = {2025 {{Topological Data Analysis}} and {{Visualization}} ({{TopoInVis}})},
	doi = {10.1109/TopoInVis68599.2025.00011},
	title = {Robust Geometric Predicates for Bivariate Computational Topology},
	year = {2025}}

@article{Sharma2025,
	author = {Sharma, Mohit and Masood, Talha Bin and List, Nanna Holmgaard and Hotz, Ingrid and Natarajan, Vijay},
	doi = {10.1109/TVCG.2025.3543619},
	journal = {IEEE Transactions on Visualization and Computer Graphics},
	pages = {1-14},
	title = {Continuous Scatterplot and Image Moments for Time-Varying Bivariate Field Analysis of Electronic Structure Evolution},
	year = {2025}}

@article{Li2025,
	author = {Mingzhe Li and Xinyuan Yan and Lin Yan and Tom Needham and Bei Wang},
	doi = {10.1109/TVCG.2025.3561300},
	journal = {{IEEE Transactions on Visualization and Computer Graphics (TVCG)}},
	number = {10},
	pages = {7951--7969},
	title = {Flexible and Probabilistic Topology Tracking With Partial Optimal Transport},
	volume = {31},
	year = {2025}}

@article{Ramamurthi2024,
	author = {Ramamurthi, Yashwanth and Chattopadhyay, Amit},
	doi = {10.1109/tvcg.2023.3314763},
	journal = {IEEE Transactions on Visualization and Computer Graphics},
	number = {9},
	pages = {5939--5952},
	title = {A Topological Distance Between Multi-Fields Based on Multi-Dimensional Persistence Diagrams},
	volume = {30},
	year = {2024}}

@article{Dobler2024,
	author = {Dobler, Alexander and N{\"o}llenburg, Martin},
	doi = {10.1111/cgf.15087},
	journal = {Computer Graphics Forum},
	number = {3},
	pages = {e15087},
	title = {Improving Temporal Treemaps by Minimizing Crossings},
	volume = {43},
	year = {2024}}

@article{Wetzels2023,
	author = {Wetzels, Florian and Pont, Mathieu and Tierny, Julien and Garth, Christoph},
	doi = {10.1109/TVCG.2023.3326601},
	journal = {IEEE Transactions on Visualization and Computer Graphics},
	number = {1},
	pages = {1095-1105},
	title = {Merge Tree Geodesics and Barycenters with Path Mappings},
	volume = {30},
	year = {2024}}

@article{Das2024,
	author = {Somenath Das and Raghavendra Sridharamurthy and Vijay Natarajan},
	doi = {10.1111/cgf.15162},
	journal = {{Computer Graphics Forum}},
	number = {6},
	title = {Time-varying Extremum Graphs},
	volume = {43},
	year = {2024}}

@inproceedings{Li2023,
	author = {Mingzhe Li and Carson Storm and Austin Yang Li and Tom Needham and Bei Wang},
	booktitle = {IEEE Visualization Conference (IEEE VIS) Short Paper},
	doi = {10.48550/arXiv.2309.04681},
	title = {Comparing Morse Complexes Using Optimal Transport: An Experimental Study},
	year = {2023}}

@inproceedings{Nilsson2023,
	author = {Nilsson, Emma and Lukasczyk, Jonas and Masood, Talha Bin and Garth, Christoph and Hotz, Ingrid},
	booktitle = {2023 Topological Data Analysis and Visualization (TopoInVis)},
	doi = {10.1109/TopoInVis60193.2023.00014},
	organization = {IEEE},
	pages = {72--81},
	title = {Probabilistic Gradient-Based Extrema Tracking},
	year = {2023}}

@article{Ramamurthi2022a,
	author = {Ramamurthi, Yashwanth and Agarwal, Tripti and Chattopadhyay, Amit},
	doi = {10.1109/tvcg.2021.3087273},
	journal = {IEEE Transactions on Visualization and Computer Graphics},
	month = dec,
	number = {12},
	pages = {4360--4374},
	publisher = {Institute of Electrical and Electronics Engineers (IEEE)},
	title = {A Topological Similarity Measure Between Multi-Resolution Reeb Spaces},
	volume = {28},
	year = {2022}}

@phdthesis{hristovHypersweepsConvectiveClouds2022,
	author = {Hristov, Petar Georgiev},
	copyright = {cc\_by\_nc\_sa\_4},
	langid = {english},
	month = jun,
	school = {University of Leeds},
	title = {Hypersweeps, {{Convective Clouds}} and {{Reeb Spaces}}},
	urldate = {2024-07-08},
	year = 2022}

@article{Wetzels2022a,
	author = {Florian Wetzels and Heike Leitte and Christoph Garth},
	doi = {doi.org/10.1111/cgf.14547},
	journal = {{Computer Graphics Forum}},
	title = {Branch Decomposition‐Independent Edit Distances for Merge Trees},
	year = {2022}}

@inproceedings{Nilsson2022Cyclone,
	author = {Nilsson, Emma and Lukasczyk, Jonas and Engelke, Wito and Masood, Talha Bin and Svensson, Gunilla and Caballero, Rodrigo and Garth, Christoph and Hotz, Ingrid},
	booktitle = {Topological Data Analysis and Visualization ({TopoInVis})},
	doi = {10.1109/TopoInVis57755.2022.00016},
	pages = {92--102},
	title = {Exploring Cyclone Evolution with Hierarchical Features},
	year = {2022}}

@article{Yan2022b,
	author = {Yan, Lin and Bin Masood, Talha and Rasheed, Farhan and Hotz, Ingrid and Wang, Bei},
	doi = {10.1109/TVCG.2022.3163349},
	journal = {IEEE Transactions on Visualization and Computer Graphics},
	pages = {1-1},
	title = {Geometry Aware Merge Tree Comparisons for Time-Varying Data with Interleaving Distances},
	year = {2022}}

@inproceedings{Rasheed2022,
	author = {Farhan Rasheed and Daniel J\"onsson and Emma Nilsson and Talha Bin Masood and Ingrid Hotz},
	booktitle = {{Topological Methods in Visualization, IEEE workshop}},
	doi = {10.1109/TopoInVis57755.2022.00018},
	title = {Subject-Specific Brain Activity Analysis in fMRI Data Using Merge Trees},
	year = {2022}}

@article{Brown2021,
	author = {Adam Brown and Omer Bobrowski and Elizabeth Munch and Bei Wang},
	doi = {10.1007/s41468-020-00063-x},
	journal = {Journal of Applied and Computational Topology},
	number = {99},
	pages = {140},
	title = {Probabilistic convergence and stability of random mapper graphs},
	volume = {5},
	year = {2021}}

@article{Yan2021b,
	author = {Lin Yan and Talha Bin Masood and Raghavendra Sridharamurthy and Farhan Rasheed and Vijay Natarajan and Ingrid Hotz and Bei Wang},
	doi = {10.1111/cgf.14331},
	journal = {{Computer Graphics Forum}},
	number = {3},
	pages = {599-633},
	title = {Scalar Field Comparison with Topological Descriptors: Properties and Applications for Scientific Visualization},
	volume = {40},
	year = {2021}}

@INPROCEEDINGS{Sharma2021,
  author={Sharma, Mohit and Masood, Talha Bin and Thygesen, Signe S. and Linares, Mathieu and Hotz, Ingrid and Natarajan, Vijay},
  booktitle={2021 IEEE Visualization Conference (VIS)}, 
  title={Segmentation Driven Peeling for Visual Analysis of Electronic Transitions}, 
  year={2021},
  volume={},
  number={},
  pages={96-100},
  doi={10.1109/VIS49827.2021.9623300}}

@inproceedings{Agarwal2021,
	address = {Cham},
	author = {Agarwal, Tripti and Chattopadhyay, Amit and Natarajan, Vijay},
	booktitle = {Topological {{Methods}} in {{Data Analysis}} and {{Visualization VI}}},
	doi = {10.1007/978-3-030-83500-2_11},
	editor = {Hotz, Ingrid and Bin Masood, Talha and Sadlo, Filip and Tierny, Julien},
	pages = {197--217},
	publisher = {Springer International Publishing},
	title = {Topological feature search in time-varying multifield data},
	year = {2021}}

@incollection{Engelke2020,
	author = {Engelke, Wito and Masood, Talha Bin and Beran, Jakob and Caballero, Rodrigo and Hotz, Ingrid},
	booktitle = {Topological Methods in Data Analysis and Visualization {VI}: Theory, Applications and Software},
	doi = {10.1007/978-3-030-83500-2_5},
	editor = {Ingrid Hotz and Talha Bin Masood and Filip Sadlo and Julien Tierny},
	publisher = {Springer},
	series = {Mathematics and Visualization},
	title = {Topology-{Based} {Feature} {Design} and {Tracking} for {Multi}-{Center} {Cyclones}},
	url = {http://arxiv.org/abs/2011.08676},
	year = {2021}}

@article{Sridharamurthy2020,
	author = {Raghavendra Sridharamurthy and Talha Bin Masood and Adhitya Kamakshidasan and Vijay Natarajan},
	doi = {10.1109/TVCG.2018.2873612},
	journal = {{IEEE Transactions on Visualization and Computer Graphics (TVCG)}},
	number = {3},
	pages = {1518-1531},
	title = {Edit Distance between Merge Trees},
	volume = {26},
	year = {2020}}

@article{Schnorr2020,
	author = {Andrea Schnorr and Dirk N. Helmrich and Dominik Denker and Torsten W. Kuhlen and Bernd Hentschesol},
	doi = {10.1109/TVCG.2018.2883630},
	journal = {{IEEE Transactions on Visualization and Computer Graphics (TVCG)}},
	number = {6},
	title = {Feature Tracking by Two-Step Optimization},
	volume = {26},
	year = {2020}}

@article{Valsangkar2019,
	author = {Valsangkar, Akash Anil and Monteiro, Joy Merwin and Narayanan, Vidya and Hotz, Ingrid and Natarajan, Vijay},
	journal = {{IEEE} Transaction on Visualization and Computer Graphics},
	number = {3},
	pages = {1460--1473},
	publisher = {IEEE},
	title = {An exploratory framework for cyclone identification and tracking},
	volume = {25},
	year = {2019}}

@article{Kopp2019,
	author = {K{\"o}pp, Wiebke and Weinkauf, Tino},
	doi = {10.1109/TVCG.2018.2865265},
	journal = {IEEE Transactions on Visualization and Computer Graphics},
	number = {1},
	pages = {534-543},
	title = {Temporal Treemaps: Static Visualization of Evolving Trees},
	volume = {25},
	year = {2019}}

@inproceedings{Soler2018,
	author = {Soler, Maxime and Plainchault, M\'{e}lanie and Conche, Bruno and Tierny, Julien},
	booktitle = {IEEE 8th Symposium on Large Data Analysis and Visualization (LDAV)},
	doi = {10.1109/LDAV.2018.8739196.},
	title = {Lifted Wasserstein matcher for fast and robust topology tracking},
	year = {2018}}

@article{Saikia2017,
	author = {H. Saikia and Tino Weinkauf},
	doi = {10.1111/cgf.13163},
	journal = {{Computer Graphics Forum}},
	number = {3},
	pages = {1--11},
	title = {Global Feature Tracking and Similarity Estimation in Time-Dependent Scalar Fields},
	volume = {36},
	year = {2017}}

@article{Tierny2017b,
	author = {Julien Tierny and Hamish Carr},
	journal = {{IEEE Transactions on Visualization and Computer Graphics (TVCG)}},
	number = {1},
	pages = {960--969},
	title = {Jacobi Fiber Surfaces for Bivariate Reeb Space Computation},
	volume = {23},
	year = {2017}}

@article{Tierny2017,
	author = {Julien Tierny and Hamish Carr},
	journal = {{IEEE Transactions on Visualization and Computer Graphics}},
	number = {1},
	pages = {960--969},
	title = {Jacobi Fiber Surfaces for Bivariate Reeb Space Computation},
	volume = {23},
	year = {2017}}

@inproceedings{lukasczyk2017nested,
	author = {Lukasczyk, Jonas and Weber, Gunther and Maciejewski, Ross and Garth, Christoph and Leitte, Heike},
	booktitle = {Computer Graphics Forum},
	number = {3},
	organization = {Wiley Online Library},
	pages = {12--22},
	title = {Nested tracking graphs},
	volume = {36},
	year = {2017}}

@inproceedings{Dey2016,
	author = {Tamal K. Dey and Facundo M\'{e}moli and Yusu Wang},
	booktitle = {Proceedings of the 27th annual {ACM-SIAM} symposium on Discrete algorithms},
	pages = {997-1013},
	title = {Mutiscale Mapper: A Framework for Topological Summarization of Data and Maps},
	year = {2016}}

@article{Carr2015,
	author = {Hamish Carr and Zhao Geng and Julien Tierny and Amit Chattopadhyay and Aron Knoll},
	journal = {{Computer Graphics Forum}},
	number = {3},
	pages = {241--250},
	title = {Fiber Surfaces: Generalizing Isosurfaces to Bivariate Data},
	volume = {34},
	year = {2015}}

@inproceedings{Chattopadhyay2014,
	author = {Amit Chattopadhyay and Hamish Carr and Duke, David and Geng, Zhao},
	booktitle = {{EuroVis} - {Short} {Papers}},
	doi = {10.2312/eurovisshort.20141156},
	editor = {Elmqvist, N. and Hlawitschka, M. and Kennedy, J.},
	publisher = {The Eurographics Association},
	title = {Extracting {Jacobi} {Structures} in {Reeb} {Spaces}},
	year = {2014}}

@article{Carr2014,
	author = {Hamish Carr and Duke, David},
	journal = {{IEEE} Transactions On Visualization And Computer Graphics},
	number = {8},
	pages = {1100--13},
	title = {Joint Contour Nets},
	volume = {20},
	year = {2014}}

@inproceedings{Widanagamaachchi2012,
	author = {Wathsala Widanagamaachchi and Cameron Christensen and Valerio Pascucci and Peer-Timo Bremer},
	booktitle = {IEEE Symposium on Large Data Analysis and Visualization (LDAV'12)},
	pages = {9--17},
	title = {Interactive exploration of large-scale time-varying data using dynamic tracking graphs},
	year = {2012}}

@article{Reininghaus2011,
	author = {Jan Reininghaus and Natallia Kotava and David G\"unther and Jens Kasten and Hans Hagen and Ingrid Hotz},
	journal = {{IEEE} Transaction on Visualization and Computer Graphics},
	number = {12},
	pages = {2045--2052},
	title = {{A Scale Space Based Persistence Measure for Critical Points in 2D Scalar Fields}},
	volume = {17},
	year = {2011}}

@inproceedings{Weber2011,
	author = {Gunther Weber and Peer-Timo Bremer and Marcus S. Day and John B. Bell and Valerio Pascucci},
	booktitle = {Topological Methods in Data Analysis and Visualization. Theory, Algorithms, and Applications. (TopoInVis'09)},
	title = {Feature Tracking Using Reeb Graphs},
	year = {2011}}

@article{Weinkauf2011,
	author = {Weinkauf, Tino and Theisel, Holger and Van Gelder, Allen and Pang, Alex},
	journal = {IEEE Transactions on Visualization and Computer Graphics},
	number = {6},
	pages = {770--780},
	publisher = {IEEE},
	title = {Stable feature flow fields},
	volume = {17},
	year = {2011}}

@article{Bremer2010,
	author = {Peer-Timo Bremer and Gunther H. Weber and Valerio Pascucci and Marcus S. Day and John B. Bell},
	journal = {IEEE Transactions on Visualization and Computer Graphics},
	number = {2},
	pages = {248--260},
	title = {Analyzing and Tracking Burning Structures in Lean Premixed Hydrogen Flames},
	volume = {16},
	year = {2010}}

@article{Cohen-Steiner2010,
	author = {David Cohen-Steiner and Herbert Edelsbrunner and John Harer and Yuriy Mileyko},
	journal = {Foundations of Computational Mathematics},
	number = {2},
	pages = {127-139},
	title = {Lipschitz Functions Have $L_p$-stable Persistence},
	volume = {10},
	year = {2010}}

@article{Bachthaler2008,
	author = {Sven Bachthaler and Daniel Weiskopf},
	journal = {IEEE Transactions on Visualization and Computer Graphics (Proceedings Visualization 2008)},
	number = {6},
	pages = {1428--1436},
	title = {Continuous Scatterplots},
	volume = {14},
	year = {2008}}

@article{Schneider2008,
	author = {Schneider, Dominic and Wiebel, Alexander and Carr, Hamish and Hlawitschka, Mario and Scheuermann, Gerik},
	doi = {10.1109/TVCG.2008.143},
	journal = {IEEE Transactions on Visualization and Computer Graphics},
	number = {6},
	pages = {1475-1482},
	title = {Interactive Comparison of Scalar Fields Based on Largest Contours with Applications to Flow Visualization},
	volume = {14},
	year = {2008}}

@inproceedings{Edelsbrunner2008,
	address = {New York, NY, USA},
	author = {Edelsbrunner, Herbert and Harer, John and Patel, Amit K.},
	booktitle = {Proceedings of the Twenty-Fourth Annual Symposium on Computational Geometry},
	doi = {10.1145/1377676.1377720},
	pages = {242--250},
	publisher = {Association for Computing Machinery},
	series = {SCG '08},
	title = {Reeb spaces of piecewise linear mappings},
	year = {2008}}

@incollection{Edelsbrunner2004b,
	author = {Herbert Edelsbrunner and John Harer},
	booktitle = {Foundations of Computational Mathematics, Minneapolis 2002},
	editor = {F. Cucker and R. DeVore and P. Olver and E. Sueli},
	pages = {37--57},
	publisher = {Cambridge Universtiy Press},
	title = {Jacobi Sets of Multiple {Morse} Functions},
	year = {2004}}

@article{martin_nto_2023,
	author = {Martin, Richard L.},
	doi = {10.1063/1.1558471},
	journal = {The Journal of Chemical Physics},
	month = {02},
	number = {11},
	pages = {4775-4777},
	title = {{Natural transition orbitals}},
	volume = {118},
	year = {2003}}

@article{Post2003,
	author = {Frits H. Post},
	journal = {Computer Graphics Forum},
	pages = {775--792},
	title = {The State of the Art in Flow Visualization: Feature Extraction and Tracking},
	volume = {22(4)},
	year = {2003}}

@inproceedings{silver1998tracking,
	author = {Silver, Deborah and Wang, Xin},
	booktitle = {Proceedings Visualization'98 (Cat. No. 98CB36276)},
	organization = {IEEE},
	pages = {79--86},
	title = {Tracking scalar features in unstructured data sets},
	year = {1998}}

@article{Samtaney1994,
	author = {Samtaney, Ravi and Silver, Deborah and Zabusky, Norman and Cao, Jim},
	journal = {Computer},
	pages = {20--27},
	publisher = {IEEE},
	title = {{Visualizing Features and Tracking their Evolution}},
	volume = {27},
	year = {1994}}

@inproceedings{Rasheed2024,
	author = {Farhan Rasheed and Abrar Naseer and Emma Nilsson and Talha Bin Masood and Ingrid Hotz},
	booktitle = {{Topological Methods in Visualization, IEEE VIS workshop}},
	doi = {10.48550/arXiv.2409.06476},
	title = {Multi-scale Cycle Tracking in Dynamic Planar Graphs},
	year = {2024}}

@article{Evers2026,
	author = {Marina Evers and Abrar Naseer and Tejas G. Murthy and Vijay Natarajan and Talha Bin Masood and Daniel Weiskopf and Ingrid Hotz},
	journal = {{Computer Graphics Forum}},
	number = {3},
	title = {Uncertainty-Aware Visual Analysis of Force Networks in 2D Granular Materials},
	volume = {45},
	year = {2026}
    }

@inproceedings{Sane2021FeatureConfidence,
booktitle = {EuroVis 2021 - Short Papers},
editor = {Agus, Marco and Garth, Christoph and Kerren, Andreas},
title = {{Visualization of Uncertain Multivariate Data via Feature Confidence Level-Sets}},
author = {Sane, Sudhanshu and Athawale, Tushar M. and Johnson, Chris R.},
year = {2021},
publisher = {The Eurographics Association},
ISBN = {978-3-03868-143-4},
DOI = {10.2312/evs.20211053}
}

@ARTICLE{Athawale2023FiberUncertainty,
  author={Athawale, Tushar M. and Johnson, Chris R. and Sane, Sudhanshu and Pugmire, David},
  journal={IEEE Transactions on Visualization and Computer Graphics}, 
  title={Fiber Uncertainty Visualization for Bivariate Data With Parametric and Nonparametric Noise Models}, 
  year={2023},
  volume={29},
  number={1},
  pages={613-623},
  doi={10.1109/TVCG.2022.3209424}}

@inproceedings{Wetzels2024ElectronDensity,
booktitle = {EuroVis 2024 - Short Papers},
editor = {Tominski, Christian and Waldner, Manuela and Wang, Bei},
title = {{Exploring Electron Density Evolution using Merge Tree Mappings}},
author = {Wetzels, Florian and Masood, Talha Bin and List, Nanna Holmgaard and Hotz, Ingrid and Garth, Christoph},
year = {2024},
publisher = {The Eurographics Association},
ISBN = {978-3-03868-251-6},
DOI = {10.2312/evs.20241069}
}

@inproceedings{Florian2024,
  author       = {Florian Wetzels and
                  Talha Bin Masood and
                  Nanna Holmgaard List and
                  Ingrid Hotz and
                  Christoph Garth},
  editor       = {Christian Tominski and
                  Manuela Waldner and
                  Bei Wang},
  title        = {Exploring Electron Density Evolution using Merge Tree Mappings},
  booktitle    = {26th Eurographics Conference on Visualization, EuroVis 2024 - Short
                  Papers, Odense, Denmark, May 27-31, 2024},
  publisher    = {Eurographics Association},
  year         = {2024},
  url          = {https://doi.org/10.2312/evs.20241069},
  doi          = {10.2312/EVS.20241069}
}

\clearpage
\includepdf[
  pages=-,
  pagecommand={},
  fitpaper=true
]{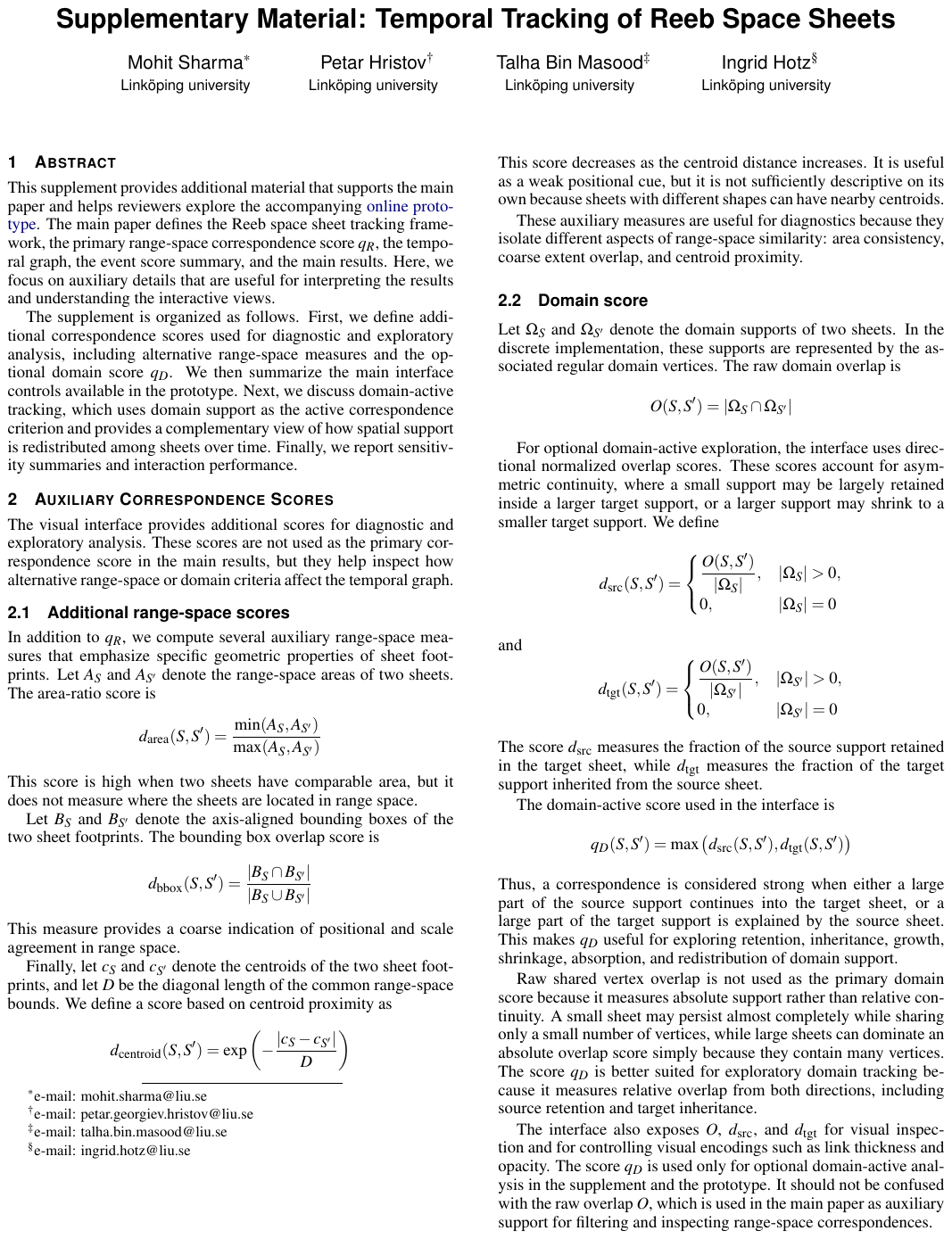}

\end{document}